\documentclass[sigconf,natbib=true]{acmart}

\AtBeginDocument{%
  \providecommand\BibTeX{{%
    \normalfont B\kern-0.5em{\scshape i\kern-0.25em b}\kern-0.8em\TeX}}}

\setcopyright{acmcopyright}
\copyrightyear{2023}
\acmYear{2023}
\acmDOI{XX.XX}

\acmConference[CIKM'23]{32nd ACM Conference on Information and Knowledge Management}{October 10--21,
  2023}{Birmingham, UK}
\begin{document}

\title{CSPM: A Contrastive Spatiotemporal Preference Model for CTR Prediction in On-Demand Food Delivery Services}


\author{ Guyu Jiang$^{\ast}$, Xiaoyun Li$^{\ast}$, Rongrong Jing, Ruoqi Zhao, Xingliang Ni$^{\dag}$, Guodong Cao, Ning Hu}
\affiliation{%
  \institution{Alibaba Group}
  \city{Beijing\&Shanghai}
  \country{China}}
\email{{guyu.qgy, xiaoyun.li, rongrongjing.jrr, ruoqi.zrq, wangzhe.nxl, guodong.cao, huning.hu}@alibaba-inc.com}

\thanks{*These authors contributed equally to this work. $^{\dag}$ Corresponding author}

\renewcommand{\shortauthors}{Guyu Jiang and Xiaoyun Li, et al.}

\begin{abstract}
Click-through rate (CTR) prediction is a crucial task in the context of an online on-demand food delivery (OFD) platform for precisely estimate the probability of a user clicking on food items. Unlike universal e-commerce platforms such as Taobao and Amazon, user behaviors and interests on the OFD platform are more location and time-sensitive due to limited delivery ranges and regional commodity supplies. However, existing CTR prediction algorithms in OFD scenarios concentrate on capturing interest from historical behavior sequences, which fails to effectively model the complex spatiotemporal information within features, leading to poor performance. To addresses this challenge, this paper introduces the \underline{C}ontrastive \underline{S}patiotemporal \underline{P}reference \underline{M}odel (CSPM), which disentangles users' spatiotemporal preferences from multiple-field features under different search states using three modules: contrastive spatiotemporal representation learning (CSRL), spatiotemporal preference extractor (StPE), and spatiotemporal information filter (StIF) . CSRL utilizes a contrastive learning framework to generate a spatiotemporal activation representation (SAR) for the search action. StPE employs SAR to activate users' diverse preferences related to location and time from the historical behavior sequence field, using a multi-head attention mechanism. StIF incorporates SAR into a gating network to automatically capture important features with latent spatiotemporal effects. Extensive experiments conducted on two large-scale industrial datasets demonstrate the state-of-the-art performance of CSPM. Notably, CSPM has been successfully deployed in Alibaba's online OFD platform Ele.me, resulting in a significant 0.88\% lift in CTR, which has substantial business implications.

\end{abstract}





\maketitle
\section{Introduction}
With the advancements in Location-Based Services (LBS) technology, on-demand food delivery (OFD) platforms like Uber Eats, Door Dash, Meituan-Dianping, and Ele.me have witnessed substantial growth in recent years, delivering millions of meals worldwide on a daily basis. Click-through rate (CTR) prediction, a crucial component of OFD platforms' advertising and recommendation systems, aims to predict the likelihood of user interaction with recommended items. For example, in Ele.me's retrieval system, users simply provide their addresses and enter food keywords, and they are immediately presented with the most relevant candidate shops and commodities for selection. Once users make their desired purchases through the payment page, timely delivery is arranged by the delivery driver.

Unlike universal e-commerce platforms such as Amazon and Taobao, user click behaviors in OFD platforms are more sensitive to factors like location and time, necessitating corresponding advertising systems to account for users' spatiotemporal preferences. Numerous methods \cite{cui2021st, lin2022spatiotemporal, wang2020calendar, du2022basm, qi2021trilateral} have been proposed for user behavior modeling, exhibiting promising experimental results. For instance, StEN \cite{lin2022spatiotemporal} explicitly models personalized spatiotemporal preferences by incorporating historical behaviors using a complex activation and attention-based network. CaledarGNN \cite{wang2020calendar} captures time structures from spatiotemporal user behaviors but overlooks the sensitive spatial needs in food delivery. BASM \cite{du2022basm} introduces a bottom-up adaptive model that leverages embedding, semantic, and classification layers to learn the spatiotemporal data distribution and enhance model fitting capacity. TRISAN \cite{qi2021trilateral} integrates the temporal relatedness of a user's historical click sequences to model the geographic closeness of items and requests based on distance or semantic similarity.

However, existing methods solely focus on capturing spatiotemporal interest from historical behavior sequences, which fails to fully exploit the spatiotemporal information contained in advertising system features. In fact, the spatiotemporal information embedded within these features is intricate in OFD scenarios. Alongside historical behavior-derived spatiotemporal interest, empirical observations for the inherent attribute in LBS systems suggest that seemingly unrelated features possess latent effects that contribute to modeling user spatiotemporal preferences. For example, while subsidy releases of red packets are generally not spatiotemporal features, they exhibit spatiotemporal characteristics in practice due to specific business strategies, attracting users who appreciate discounts. Moreover, spatiotemporal preferences can be influenced by the current search state, including the query, user location, and time information, further increasing the modeling challenge. For instance, when users search for "beef", they may prefer to purchase raw food for cooking if they are at home, whereas fast food may be preferred if they are at work.

To address these challenges, we propose the Contrastive Spatiotemporal Preference Model (CSPM) that considers both explicit spatiotemporal preferences relevant to users' historical behaviors and implicit spatiotemporal preferences contained in context and item features. CSPM utilizes the contrastive spatiotemporal representation learning (CSRL) module to transform user search behavior, including the query, time, and location features, into a spatiotemporal activation representation (SAR) that will be used to model the subsequent spatiotemporal preferences. To comprehensively capture the spatiotemporal preference patterns embedded in multi-field features, CSPM introduces the spatiotemporal preference extractor (StPE) module and the spatiotemporal information filter (StIF) module. StPE module employs a multi-head attention mechanism and SAR from CSRL to extract explicit spatiotemporal preferences correlated with users' historical behaviors. StIF module incorporates SAR into a gating network to automatically capture important features with latent spatiotemporal effects. By jointly optimizing the CSRL triplet loss and the CTR prediction cross-entropy loss, CSPM learns a comprehensive spatiotemporal preference model to enhance CTR prediction accuracy. The contributions of this paper can be summarized as follows:


\begin{itemize}
    \item We propose the contrastive spatiotemporal preference model, which utilizes contrastive spatiotemporal representation learning to generate an activation signal for the current search state, enabling the modeling of explicit and implicit spatiotemporal preferences from multi-field features.
    \item We develop two user preference model modules: the spatiotemporal preference extractor, which extracts explicit spatiotemporal preferences correlated with users' historical behaviors, and the spatiotemporal information filter, which activates implicit spatiotemporal information contained in context and item features.
    \item We evaluate CSPM on two large-scale real-world datasets, and the experimental results demonstrate its significant outperformance compared to other state-of-the-art methods in CTR prediction. Additionally, we release an industrial dataset for the OFD industry to address the scarcity of public datasets in this domain.
\end{itemize}

\begin{figure*}[htbp]
  \centering
  \includegraphics[width=\linewidth]{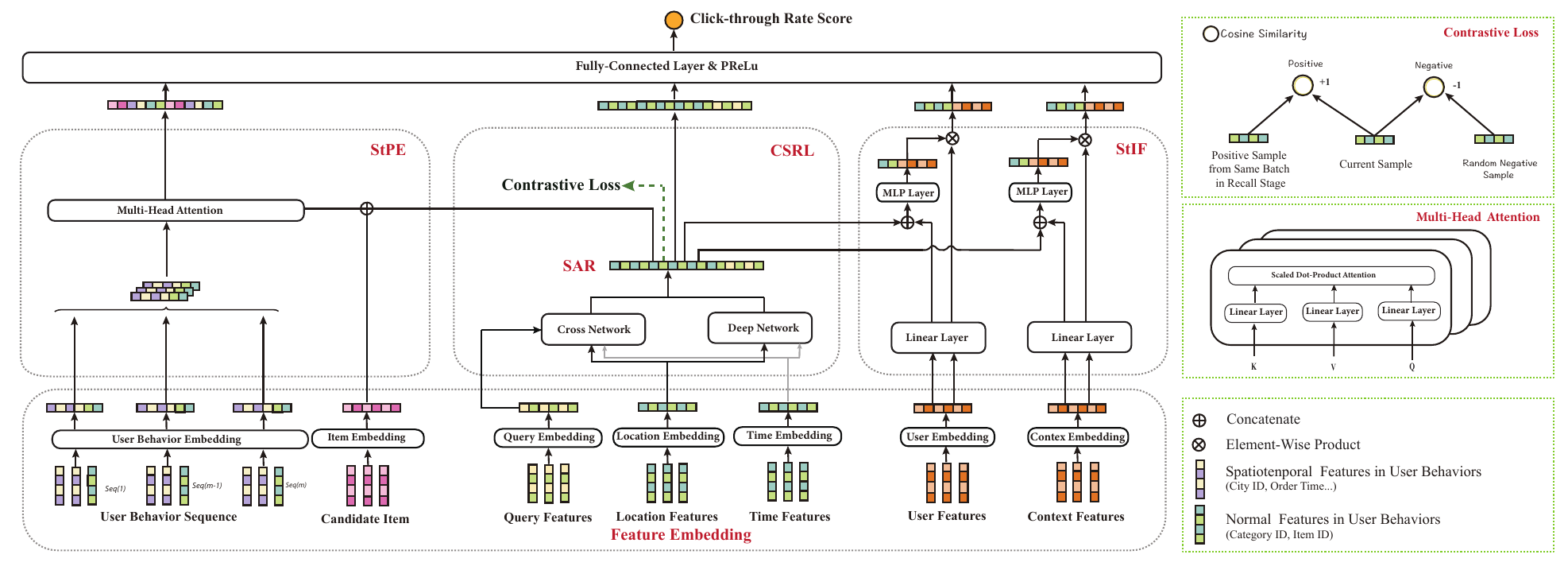}
  \caption{The framework of our proposed CSPM. CSPM contains three main modules: CSRL, StPE, and StIF.}
  \label{fig:cspm3}
  \Description{xx}
\end{figure*}




\section{Methodology}

\subsection{Preliminary}
Considering a training dataset $\mathcal{D}=\{(x_i,y_i )\}_{i=1}^{|\mathcal{D}|}$ , where $x_i$ and $y_i$ represent the feature set and binary click label of the $i_{th}$ sample. In an OFD recommendation system, the feature set usually can typically be divided into several fields, denoted as $x_i=(x_i^e),e\in\{E_S, E_I, E_Q, E_L, E_T, E_U, E_C\}$. These superscripts correspond to different fields, including user behavior sequence, candidate item from the recall stage, search query, search location, search time, user, and context. We formulate the task of CTR prediction as a learning problem, where the goal is to estimate the probability 
\begin{equation}
\hat{y}=\mathbb{P}(y=1|x,\mathcal{D}) = f(x).
\end{equation}
Here, $f(\cdot)$ represents a learnable function, and $\hat{y}\in[0,1]$ represents the predicted probability of the user clicking on the target item. To compute the prediction error of CTR, we employ the cross-entropy loss given by:
\vspace{-0.2cm}
\begin{equation}
\mathcal{L} _{CTR}=-\frac{1}{\left| \mathcal{D} \right|}\sum_{i=1}^{\left| \mathcal{D} \right|}{y_i\log \hat{y}_i+\left( 1-y_i \right) \log \left( 1-\hat{y}_i \right)}.
\end{equation}
\subsection{Contrastive Spatiotemporal Representation Learning}
The user's search action under different spatiotemporal contexts, could imply distinct needs and provides valuable insights into extracting historical and latent spatiotemporal preferences. In this paper, we utilize query, location, and time features $x^e$ $(e\in\{E_Q,E_L,\\E_T\})$ to describe the search state or action. To enhance the distinguishability of the search state and activate subsequent spatiotemporal preferences effectively, we introduce a CSRL module. This module assumes that representations of semantically similar search queries under close geohashes and within the same time period should be close together in the representation space, while dissimilar search states should be far apart. CSRL transforms these search state features into low-dimensional, dense, continuous vectors using embedding techniques and utilizes a DCN-v2 \cite{dcnv2} to generate a high-order cross-field vector $s$, called spatiotemporal activation representation (SAR). For constructing positive and negative pairs, we select locations from the same batch of recall results and assign temporal proximity by considering a half-hour interval ahead or behind the search time. Positive pairs are formed with these spatiotemporal constraints, while negative pairs are randomly selected. 
The contrastive loss in CSRL adopts the triplet loss as follows:
\begin{equation}
\begin{aligned}
    \mathcal{L}_{CL}=-\frac{1}{|\mathcal{D}|}\sum_{i=1}^{|\mathcal{D}|}&\sum_{j=1}^{n_v} \max(\cos(s_i^p,s_i)-\cos(s_i^{n_j},s_i)+m,0),
\end{aligned}
\end{equation}
where $n_v$ represents the number of randomly selected negative samples, $\cos(\cdot \hspace{0.2 em},\cdot)$ calculates the cosine similarity between the two vectors, $m$ is a margin, and $s_i^p$ and $s_i^{n_j}$ denote the SARs of positive and negative samples, respectively.

\subsection{Spatiotemporal Preference Extractor}
The behavior sequence records a user's combination of purchase incidents within a specific time period. We argue that the current search action plays a pivotal role in eliciting a user's historical preference. Therefore, we propose the StPE module, which leverages the SAR of search state from CSRL to query the behavior sequence. Drawing on the ideas of \cite{song2019autoint}, the backbone of StPE is a multi-head attention mechanism \cite{vaswani2017attention} that concatenates the spatiotemporal activation representation and the candidate item embedding as the activation query. The user behavior sequence is regard as the key and value by different linear transformations. The formulation of StPE can be expressed as follows:
\begin{equation}
u_h={\rm{Attention}}\left(\left[v^{E_C};s\right]W_h^Q,U^{E_S} W_h^K,U^{E_S} W_h^V\right),
\end{equation}
where $v^{E_C}$ and $U^{E_S}$ denote the candidate item-field representation and the user sequence-field representation matrix, respectively. $\rm Attention(\cdot)$ represents the self-attention mechanism, $[\cdot \hspace{0.2 em},\cdot]$ is the concatenation operator, and $W_h^Q$, $W_h^K$, and $W_h^V$ are the parameters of the $h_{th}$ head.

\subsection{Spatiotemporal Information Filter}

In addition to capturing the explicit spatiotemporal preference activated by the current search action, we further introduce the StIF module, which utilizes the search state representation to extract as much implicit spatiotemporal information as possible from the features. In this paper, StIF automatically identifies significant features with latent spatiotemporal effects from the user and context fields, depending on the feature construction of the specific recommendation system. To achieve this, we adopt the gating method GateNet \cite{huang2020gatenet}, which incorporates a learnable feature gating unit to effectively capture high-order interactions. The core idea can be described as follows:
\vspace{-0.2cm}

\begin{equation}
    o_i=s_i  \otimes \sum_j w_{i,j},
\end{equation}
\begin{equation}
    w_{i,j}=\Phi_w (s_i,z_j ),
\end{equation}
where $s_i$ represents the spatiotemporal activation representation of the current search state, $z_j$ denotes the features from the $E_U$ and $E_C$ fields, and $\Phi_w$ is an MLP layer and sigmoid function that generate feature importance weights through information interaction.

The final loss function is a summation of the CTR prediction loss and the spatiotemporal contrastive loss:
\begin{equation}
    \mathcal{L}_{total}=\alpha \mathcal{L}_{CTR}+(1-\alpha)\mathcal{L}_{CL},
\end{equation}
where $\alpha\in[0,1]$  is a hyperparameter used to balance the contributions of $\mathcal{L}_{CTR}$ and $\mathcal{L}_{CL}$.

\section{EXPERIMENTS}

\subsection{Experimental Setting}
\subsubsection{Datasets.}
The \textbf{Real-OFD Public Dataset\footnote{https://tianchi.aliyun.com/dataset/dataDetail?dataId=120281}} is randomly collected from the retrieval logs of active users on the Ele.me OFD platform, comprising approximately 400,000 samples. It encompasses more than 30 features, including user behavior sequences spanning a maximum length of 50, recorded over a month. \noindent The \textbf{Industrial Dataset} is constructed using impression and click logs obtained from the Ele.me OFD platform. The training set consists of samples from the preceding 89 days, while the test set comprises data from the subsequent day. The user's click behavior within the past 45 days is considered as the historical behavior sequence.
\subsubsection{Baselines.}
We compare six widely-used and industry-proven CTR models: Wide \& Deep \cite{cheng2016wide}, DCN \cite{wang2017deep}, DeepFM \cite{guo2017deepfm}, DIN \cite{zhou2018deep}, DIEN \cite{zhou2019deep}, and AutoInt \cite{song2019autoint}. Additionally, to demonstrate the effectiveness of our proposed modules in CSPM, we apply the CSRL module to DIN and activate other features using StPE and StIF modules. Furthermore, to compare the learning ability for spatiotemporal features, we provide comparisons of three spatiotemporal CTR models: StEN \cite{lin2022spatiotemporal}, BASM \cite{du2022basm}, and TRISAN \cite{qi2021trilateral}.

All models in this paper are trained using the same training strategy and features. We utilize AdagradDecay \cite{duchi2011adaptive} as the optimizer for these datasets and apply exponential decay, starting with a learning rate of 0.01 for the Real-OFD Public Dataset and 0.001 for the Industrial Dataset. We set the feature embedding dimension to 32 and truncate the length of users' behaviors to 100.

For evaluating the model performance, we employ the widely-used metric AUC (Area Under ROC), which has been validated as a good measurement for CTR prediction \cite{covington2016deep}. AUC is insensitive to the classification threshold and the positive ratio, with a maximum value of 1 indicating best performance.

\subsection{Performance Comparison} 
In this subsection, we compare CSPM with three types of methods. It is worth noting that a 0.1\% improvement in AUC is considered quite significant on an online e-commerce platform.

As shown in Table \ref{tab:table1}, CSPM exhibits better performance compared to five mainstream CTR models. We observe that DIN, as one of the most commonly-used models in online systems, shows excellent performance on both datasets due to its ability to capture users' interests using the attention mechanism. However, the performance of other methods varies widely on both datasets. Besides, to investigate the generalization of our proposed modules, we conduct variation experiments by applying the CSRL, StPE, and StIF modules to DIN. As seen in Table \ref{tab:table1}, these modules contribute to improving the model's effectiveness. Specifically, the CSRL module shows a more significant improvement compared to the other two modules, indicating the importaness of the search state's spatiotemporal representation. It is noteworthy that DIN with all the modules achieves a performance similar to that of CSPM. Such result further demonstrates that our proposed spatiotemporal modules can be applied to existing models to capture users' spatiotemporal preferences. In the third group of experiments, we compare CSPM to three spatiotemporal-enhanced models. StEN and BASM outperform TRISAN, which lacks exploration of users' spatiotemporal behaviors. In summary, CSPM excels in capturing users' strong spatial and temporal preferences, thereby improving the performance of models.

\begin{table}
  \caption{Model Performance (AUC) on the Industrial Dataset and Real-OFD Dataset}
  \label{tab:table1}
  \begin{tabular}{lcc}
    \toprule
    Model & Industrial & Real-OFD \\
    \midrule
        Wide \& Deep \cite{cheng2016wide}	 &0.8570	&0.7446\\
        DCN \cite{wang2017deep} & 0.8568 	& 0.7449\\
        DeepFM \cite{guo2017deepfm} 
        & 0.8576	&0.7450 \\
        DIN	\cite{zhou2018deep} &0.8581	    &0.7452\\
        DIEN\cite{zhou2019deep}	&0.8561	    &0.7442\\
        AutoInt \cite{song2019autoint} &0.8574	&0.7445\\
        \hline
        DIN+CSRL    &0.8589	       &0.7459\\
        DIN+StPE+StIF &0.8585	       &0.7453\\
        DIN+CSRL+StPE &0.8585	    &0.7467\\
        DIN+CSRL+StIF &0.8592	    &0.7466\\
        DIN+CSRL+StPE+StIF &0.8601	&0.7479\\
        \hline
        StEN\cite{lin2022spatiotemporal} 	                         &0.8596   &0.7469\\
        BASM\cite{du2022basm}           &0.8586	  &0.7471\\
        TRISAN\cite{qi2021trilateral}   &0.8571	  &0.7459\\
         \hline
         CSPM	       &0.8605	    &0.7481\\
  \bottomrule
\end{tabular}
\end{table}

\subsection{Ablation Studies}

To validate the effectiveness of our CSPM, we conducted ablation studies. Firstly, we examined the effect of our CSRL module by removing the contrastive loss and the DCN-v2 module, respectively. As shown in Table \ref{tab:table2}, the experimental results demonstrate the effectiveness of these two modules, particularly the contrastive loss, which facilitates the learning of valuable spatiotemporal representations. Although the StPE module outperforms the StIF module by 0.23\% on the industrial dataset and 0.18\% on the Real-OFD dataset, indicating that the effect of explicit spatiotemporal preferences is more noticeable than implicit spatiotemporal preferences, it is also evident that StIF, activated by CSRL, contributes to performance improvement, thereby confirming the effectiveness of capturing latent spatiotemporal effects.

\begin{table}
  \caption{Ablation Performance (AUC) the Industrial Dataset and Real-OFD Dataset.}
  \label{tab:table2}
  \begin{tabular}{lcc}
    \toprule
    Model & Industrial & Real-OFD \\
    \midrule
    w/o CSRL ($\mathcal{L}_{CL}$)  & 0.8557  & 0.7451\\
    w/o CSRL (DCN-v2)    & 0.8569  & 0.7458\\
    w/o CSRL          & 0.8556  & 0.7425\\
    w/o StPE	          & 0.8573  & 0.7461\\
    w/o StIF	          & 0.8596  & 0.7479\\
    w/o StPE+StIF	      & 0.8561  & 0.7439\\
    w/o CSRL+StPE+StIF  & 0.8550  & 0.7412\\
    \hline
    CSPM              & 0.8605  & 0.7481\\
  \bottomrule
\end{tabular}
\end{table}

\subsection{Result from online A/B testing}
\begin{figure}[htbp]
  \centering
  \includegraphics[width=0.48\textwidth]{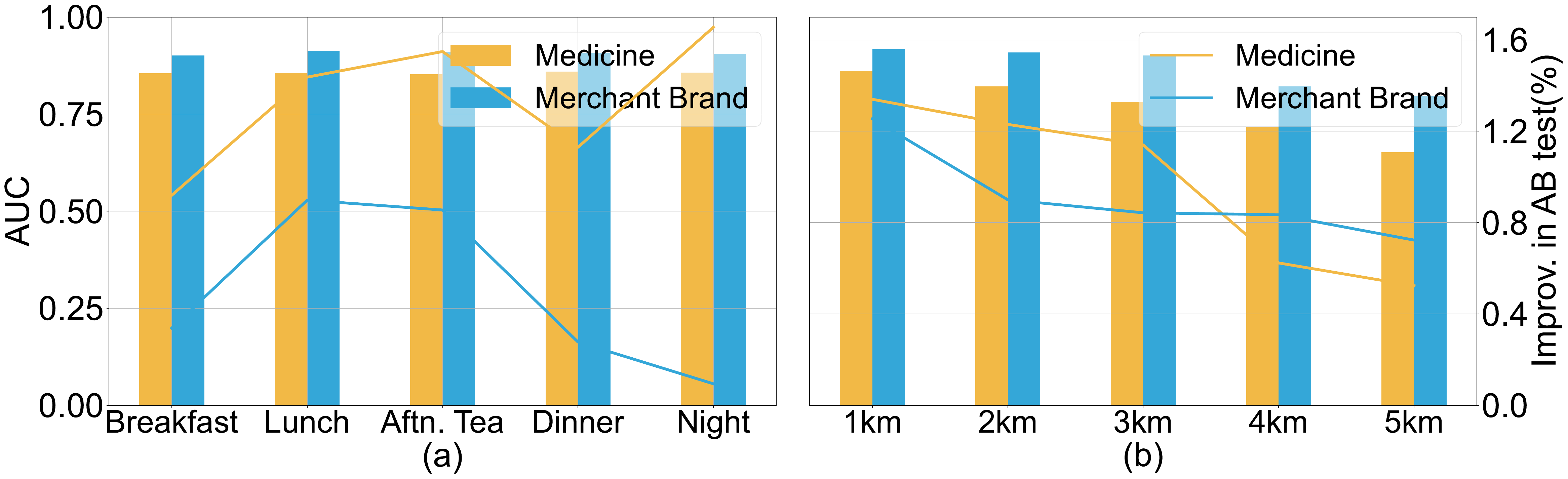}
  \caption{
  The different improvements achieved by CSPM.
  }
  \label{fig:ab}
  \Description{xx}
\end{figure}
From January 20, 2023, to February 18, 2023, we conducted online A/B testing on the Ele.me OFD platform. The A/B tests were carried out using two buckets: one as the experimental group and the other as the baseline group. To maximize the confidence level of the A/B experiment, we randomly assigned users to the buckets. After one month of cumulative testing, each test bucket accumulated close to twenty million data points. Comparing CSPM with the online-serving base model, we observed an improvement in CTR by 0.88\% and a 1.0\% increase in the purchase rate of each impression. These improvements were verified through hypothesis testing. Fig \ref{fig:ab} illustrates the different degrees of improvements achieved by CSPM under medicine and merchant brand two types of queries in all periods and delivery distances. 
CSPM has made significant advancements in the service requests that require spatiotemporal sensitivity, such as medical queries during office/night hours and within a 3 km distance. Currently, CSPM has been deployed online and serves the main traffic, resulting in a significant enhancement in user experience and a substantial growth in business revenue.

\section{CONCLUSIONS}
This paper presents CSPM, a solution for tackling the challenge of diverse user behavior and interests in different locations and times, which affects CTR prediction in OFD platforms. CSPM comprises CSRL, StPE, and StIF modules to model complex spatiotemporal preferences within multi-field features. CSRL transforms the current user search state into a spatiotemporal activation representation for subsequent spatiotemporal preferences. StPE, utilizing a multi-head attention mechanism and SAR, extracts diverse location and time preferences from the historical behavior sequence field. StIF incorporates a gating network to capture important features with latent spatiotemporal effects from user and context fields.  These modules disentangle users' spatiotemporal preferences under different search states from multiple-field features. Extensive experiments on two large-scale industrial datasets demonstrate the superior performance of CSPM. A/B testing and online deployment confirm significant business improvement with CSPM.

\bibliographystyle{ACM-Reference-Format}
\bibliography{sample-base}

\appendix

\end{document}